# Scientific Workflow Systems for 21st Century, New Bottle or New Wine?

## Invited Short Paper

[1]Yong Zhao, [2]Ioan Raicu, [2,3,4]Ian Foster
[1]Microsoft Corporation, Redmond, WA, USA
[2] Department of Computer Science, University of Chicago, Chicago, IL, USA
[3]Computation Institute, University of Chicago, Chicago, IL, USA
[4]Math & Computer Science Division, Argonne National Laboratory, Argonne, IL, USA
yozha@microsoft.com, iraicu@cs.uchicago.edu, foster@mcs.anl.gov

## Abstract

*With the advances in e-Sciences and the growing complexity of scientific analyses, more and more scientists and researchers are relying on workflow systems for process coordination, derivation automation, provenance tracking, and bookkeeping. While workflow systems have been in use for decades, it is unclear whether scientific workflows can or even should build on existing workflow technologies, or they require fundamentally new approaches. In this paper, we analyze the status and challenges of scientific workflows, investigate both existing technologies and emerging languages, platforms and systems, and identify the key challenges that must be addressed by workflow systems for e-science in the 21$^{st}$ century.*

## 1. Introduction

Scientific workflow has become increasingly popular in modern scientific computation as more and more scientists and researchers are relying on workflow systems to conduct their daily science analysis and discovery. With technology advances in both scientific instrumentation and simulation, the amount of scientific datasets is growing exponentially each year, such large data size combined with growing complexity of data analysis procedures and algorithms have rendered traditional manual processing and exploration unfavorable as compared with modern *in silico* processes automated by scientific workflow systems (SWFS). While the term workflow speaks of different things in different context, we find in general SWFS are engaged and applied to the following aspects of scientific computations: 1) *describing complex scientific procedures,* 2) *automating data derivation processes*, 3) *high performance computing (HPC) to improve throughput and performance*, and 4) *provenance management and query*.

Workflows are not a new concept and have been around for decades. There were a number of coordination languages and systems developed in the 80s and 90s [1,7], which share many common characteristic with workflow systems (i.e. they describe individual computation components and their ports and channels, and the data and event flow between them). They also coordinate the execution of the components, often on parallel computing resources. Furthermore, business process management systems have been developed and invested in for years; there are many mature commercial products and industry standards such as BPEL [2]. In the scientific community there are also many emerging systems for scientific programming and computation [5,22]. Before we jump on developing yet another workflow system, a fundamental question to ask is whether we can use existing technologies, or we should invent new languages and systems in order to achieve the four aspects mentioned earlier that are essential to scientific workflow systems. This paper identifies the challenges to workflow development in the context of scientific computation; we present an overview of some of the existing technologies and emerging systems, and discuss opportunities in addressing these challenges.

## 2. Multi-core processor architectures

Software development has been on a free ride for performance gain as chipmakers continue to follow Moore's Law in doubling up transistors in minuscule space. Little consideration has been given to code parallelization since it has not been essential for the average computer user until recently, when single CPU core performance growth stagnated and multi-core processors emerged on the market in 2005.

Due to the limitations to effectively increasing processor clock frequency, hardware manufactures started to physically reorganize chips into what we call the multi-core architecture [10], involving linking several microprocessor cores together on the same semiconductor. Various manufactures from Intel, AMD, IBM, Sun, have released dual-core, quad-core, eight-core, and 64-threaded processors in the past few years [13,21]. Given that 128-threaded SMP systems are a reality today [21], it is reasonable to assume that 1024 CPU cores/threads or more per SMP system will be available in the next decade.

The new multi-core architecture will force radical changes in software design and development. We are already seeing significant increase of research interests in concurrency and parallelism, and multi-core software development. The number of multiprocessor research papers has increased sharply since year 2001, surpassing the peak point in all the past years [10]. Concurrency is one of the next big challenges in how we write software simply because our industry has been driven by requirements to write ever larger systems that solve ever more complicated problems and exploit the ever greater computing and storage resources that are available [18].

## 3. The data deluge challenge in science

Within the science domain, the data that needs to be processed generally grows faster than computational resources and their speed. The scientific community is facing an imminent flood of data expected from the next generation of experiments, simulations, sensors and satellites. Scientists are now attempting calculations requiring orders of magnitude more computing and communication than was possible only a few years ago. Moreover, in many currently planned and future experiments, they are also planning to generate several orders of magnitude more data than has been collected in the entire human history [9].

For instance, in the astronomy domain the Sloan Digital Sky Survey (http://www.sdss.org) has datasets that exceed 10 terabytes in size. They can reach up to 100 terabytes or even petabytes if we consider multiple surveys and the time dimension. In physics, the CMS detector being built to run at CERN's Large Hadron Collider (http://lhc.web.cern.ch/lhc) is expected to generate over a petabyte of data per year. In the bioinformatics domain, the rate of growth of DNA databases such as GenBank (http://www.psc.edu/general/software/packages/genbank/) and EMBL (European Molecular Biology Laboratory, http://www.embl.org) has been following an exponential trend, with a doubling time estimated to be 9-12 months.

To enable the storage and analysis of large quantities of data and to achieve rapid turnaround, data needs to be distributed over thousands to tens of thousands of compute nodes. In such circumstances, data locality is crucial to the successful and efficient use of large scale distributed systems for data-intensive applications [19]. Scientific workflows are generally executed on a shared infrastructure such as TeraGrid (http://www.teragrid.org), Open Science Grid (http://www.opensciencegrid.org), and dedicated clusters, where data movement relies on shared file systems that are known bottlenecks for data intensive operations. If data analysis workloads have locality of reference, then it is feasible to cache and replicate data at each individual compute node, as high initial data movement costs can be offset by many subsequent data operations performed on cached data [15].

Modern scientific workflow systems need to set large scale data management as one of its primary objectives, and to ensure data movement is minimized by intelligent data-aware scheduling both among distributed computing sites (assuming that each site has a local area network shared storage infrastructure), and among compute nodes (assuming that data can be stored on compute nodes' local disk and/or memory).

## 4. Supercomputing vs. Grid Computing

Supercomputers had their golden age back in the 80s when there were virtually no other choices in dealing with compute-intensive tasks. They were applied mostly to scientific modeling and simulation in various disciplines such as high energy physics, earth science, biology, mechanical engineering etc. Some typical applications included weather forecasting, missile trajectory simulation, airplane wind tunnel simulation, genomics etc. However, supercomputers are expensive and scarce resources where only national laboratories, government agencies and some universities have access to them; and the parallel architectures of supercomputers often dictate the use of special programming techniques to exploit their speed, such as special-purposed FORTRAN compilers, PVM, MPI and OpenMP [9].

Over the last decade, we have observed processor speeds, storage capacity per drive, and network bandwidth increase 100~1000 times. As a consequence, cluster computing and Grid computing environments that leverage the cheaper commodity computing and storage hardware have been actively adopted for scientific computations. Cluster computing usually involves homogeneous machines interconnected by high speed network with locally accessible storage in one administrative domain, where Grid computing focuses on distributed resource sharing and coordination across multiple "virtual organizations" that may span many geographically distributed administrative domains. Grids can also be categorized into Computational Grids and Data Grids, where the former mostly tackle computation intensive tasks, and the latter target data-intensive sciences.

With the introduction of multi-core architectures, the separation between Grid Computing and Supercomputing is becoming less clear. Many supercomputers are being built on multi-core chips with high speed interconnection. The Cray XT5 system (http://www.cray.com/products/xt5/index.html) uses thousands commodity Quad-Core AMD Opteron™ processors and has a unified Linux environment. The latest IBM BlueGene/P Supercomputer (BG/P, http://www.research.ibm.com/bluegene/) has quad core processors with a total of 160K-cores, and has support for a lightweight Linux kernel on the compute nodes, making it significantly more accessible to new applications [17]. Finally, a smaller system named SiCortex (http://www.sicortex.com/) is also worth mentioning; it boasts 6-core processors for a total of 5832-cores, and runs a standard Linux environment.

Supercomputers (e.g. IBM BlueGene) have traditionally been designed and used for tightly coupled massively parallel applications, typically implemented in MPI. They have not been an ideal preferred platform for executing loosely coupled applications that are typical in many scientific workflows. Grids have seen success in the execution of tightly coupled parallel applications, but they has been the platform of choice for loosely coupled applications mostly due to the flexibility and granularity of the resource management and the execution of single processor jobs with ease. Work is underway within both the Falkon [14] and Condor [20] projects to enable the latest BG/P to efficiently support loosely coupled serial jobs without any modifications to the respective applications, and hence enabling an entirely new class of applications that were never candidates as possible use cases for the BlueGene/P supercomputer.

Scalability and performance are top priorities for SWFS. To this end, it is necessary to leverage supercomputing resources as well as Grid computing infrastructures for large scale parallel computations.

## 5. Existing and emerging workflow technologies

DAGMan (http://www.cs.wisc.edu/condor/dagman) and Pegasus [6] are two systems that are commonly referred to as workflow systems and have been widely applied in Grid environments. DAGMan provides a workflow engine that manages Condor jobs organized as directed acyclic graphs (DAGs) in which each edge corresponds to an explicit task precedence. Both systems focus on the scheduling and execution of long running jobs.

Taverna [12] is an open source workflow system particularly focused on bioinformatics applications and services, and it is based on the XScufl (XML Simple Conceptual Unified Flow) language. Kepler [11] is a scientific workflow system that builds on the Ptolemy-II system (http://ptolemy.eecs.berkeley.edu/ptolemyII/), which is a visual modeling tool written in Java. Triana [4] is a GUI-based workflow system for coordinating and executing a collection of services. All these systems have some visual interfaces (also referred to as workbenches) that allow the graphical composition of workflows.

While all of the existing SWFS possess great features and address many aspects of workflow specification, execution and management problems, it is unrealistic to expect one system to cover all the bases. The Workflow Bus project [23] instead tries to leverage multiple existing workflow systems to compliment each other in implementing aggregated functions and services.

Finally, the evolutions of workflows themselves (explorations) are vital in scientific analysis. VisTrails [3] captures the notion of an evolving dataflow, and implements a history management mechanism to maintain versions of a dataflow, thus allowing a scientist to return to previous steps, apply a dataflow instance to different input data, explore the parameter space of the dataflow, and (while performing these steps) compare the associated visualization results.

In response to the pressing demand of scientific applications, and the hunger for computing power, there have been a few emerging languages and systems that try to tackle the problems taking unconventional approaches.

MapReduce [5] is regarded as a power-leveler that solves complicated computation problems using brutal-force computation power. It provides a very simple programming model and powerful runtime system for the processing of large datasets. The programming model is based on just two key functions: "map" and "reduce," borrowed from functional languages. The MapReduce runtime system automatically partitions input data and schedules the execution of programs in a large cluster of commodity machines. The system is made fault tolerant by checking worker nodes periodically and reassigning failed jobs to other worker nodes. MapReduce has been mostly applied to document processing problems, such as distributed indexing, sorting, and clustering.

The Fortress language (http://fortress.sunsource.net) recently released by Sun Microsystems is a new programming language designed for HPC, and aims to improve programmability and productivity in scientific computation. The language has been designed from ground up, supporting mathematical notation (in Unicode) and physical units and dimensions, static type checking of multidimensional arrays and matrices, and rich functionality in libraries. It supports transactions, specification of locality, and implicit parallel computation (e.g. parallel for loops). Although Fortress in a strict sense is not a workflow language, and its adoption remains to be seen, it provides the higher level abstractions and functionalities for building a parallel workflow language.

Microsoft Windows Workflow Foundation (WWF) [16] provides a generic framework for workflow development and execution. It is focused on integrating diverse components within an application, allowing a workflow to be deployed and managed as a native part of the application. The fundamental idea behind WWF is that each activity is modeled as a resumable program statement, and the invocation of an activity is asynchronously organized, thus a program can be compared to a bookmark, which can be frozen in action, serialized into persistent storage, and resumed after arbitrarily long time later. However, WWF is not a full-fledged workflow management system in that it lacks administration, monitoring, retry mechanism, load balancing, etc. for a production environment.

Star-P (http://www.interactivesupercomputing.com) approaches the integration of scientific applications and HPC via language extension – allowing scientists to work in their familiar programming environments such as MATLAB, Python, and R, with some parallel directives. Internally the system can schedule the execution of parallel tasks to a computation cluster pre-configured with scientific calculation libraries. The system has been applied to a wide variety of computation problems, but the performance improvement is mostly intra-application parallelization, instead of inter-component coordination and management.

Swift [22] is an emerging system that bridges scientific workflows with parallel computing. It is a parallel programming tool for rapid and reliable specification, execution, and management of large-scale science and engineering workflows. Swift takes a structured approach to workflow specification, scheduling and execution. It consists of a simple scripting language called SwiftScript for concise specifications of complex parallel computations based on dataset typing and iterations, and dynamic dataset mappings for accessing large scale datasets represented in diverse data formats. The runtime system relies on the CoG Karajan workflow engine for efficient scheduling and load balancing, and it integrates the Falkon [14] light-weight task execution service for optimized task throughput and resource efficiency delivered by a streamlined dispatcher, a dynamic resource provisioner, and the data diffusion mechanism to cache datasets in local disk or memory and dispatch tasks according to data locality.

## 6. Call for scientific workflow systems

Existing technologies and systems already address many of the fundamental issues in scientific workflow specification and management, and many of them have been successful applied to various scientific applications across multiple science disciplines. However, modern multi-core architectures and parallel and distributed computing technologies, and the exponentially growing scientific data are bound to change the landscape and evolution of scientific workflow systems. As already being manifested by the few emerging systems, the science community is demanding both specialized, domain-specific languages to improve productivity and efficiency in writing concurrent programs and coordination tools, and generic platforms and infrastructures for the execution and management of large scale scientific applications, where scalability and performance are major concerns. High performance computing support has become a indispensable piece of such workflow languages and systems, as there is no other viable way to get around the large storage and computing problems emerging in every discipline of $21^{st}$ century e-science, although what may be the best approach to enabling scientists to leverage HPC technologies as transparent and efficient as possible remains unanswered.

In the science domain, there is an increasing need for programming languages to expose parallelism, whether it's done explicitly or implicitly, to specify the concurrency within a component, or across multiple independent components. There is a need for new parallel or workflow languages that adopt implicit parallelism where data dependencies can be discovered by its compiler, and independent tasks in the orders of hundreds of thousands can be scheduled to run in clustered or Grid environments. Such systems could achieve improvements in both manageability and productivity.

Scientific workflow systems aim to provide a simple concise notation that allows easy parallelization and supports the composition of large numbers of parallel computations, therefore they may not need all the constructs and features in a full-fledged conventional language, and implicit parallelism is preferred to explicit parallelism specification, as the latter requires expertise and attention to the details of parallel programming, which may be difficult for end users. But in the mean time sometimes scientists do need more control in specifying how to distribute their applications and datasets.

We are also in need of common generic infrastructures and platforms in the science domain for workflow administration, scheduling, execution, monitoring, provenance tracking etc. While business process management has industry agreed upon standards and steering committees, we don't have these in the science domain, where often time people reinvent the wheel in developing their in-house yet another SWFS, and there is no easy way in integrating various workflow systems and specifications. We also argue that in order to address all the important issues such as scalability, reliability, scheduling and monitoring, data management, collaboration, workflow provenance, and workflow evolution, one system cannot fit all needs. A structured infrastructure that separates the concerns of workflow specification, scheduling, execution etc, yet is organized on top of components that specialize on one or more of the areas would be more appropriate.

## 7. References

[1] Ahuja, S., Carriero, N., and Gelernter, D., "Linda and Friends", IEEE Computer 19 (8), 1986, pp. 26-34.
[2] Andrews, T., Curbera, F., Dholakia, H., Goland, Y., Klein, J., Leymann, F., Liu, K., Roller, D., Smith, D., Thatte, S., Trickovic, I., Weerawarana, S.: Business Process Execution Language for Web Services, Version 1.1. Specification, BEA Systems, IBM Corp., Microsoft Corp., SAP AG, Siebel Systems (2003).
[3] Callahan, S.P., Freire, J., Santos, E., Scheidegger, C.E., Silva C.T. and Vo, H.T. Managing the Evolution of Dataflows with VisTrails. IEEE Workshop on Workflow and Data Flow for Scientific Applications (SciFlow) 2006.
[4] Churches, D., Gombas, G., Harrison, A., Maassen, J., Robinson, C., Shields, M., Taylor, I., Wang, I., Programming Scientific and Distributed Workflow with Triana Services. Concurrency and Computation: Practice and Experience. Special Issue on Scientific Workflows, 2005.
[5] Dean, J. and Ghemawat, S., MapReduce: Simplified data processing on large clusters. OSDI, 2004.
[6] Deelman, E., Singh, G., Su, M.-H., Blythe, J., Gil, Y., Kesselman, C., Mehta, G., Vahi, K., Berriman, G.B., Good, J., Laity, A., Jacob, J.C. and Katz, D.S. Pegasus: A Framework for Mapping Complex Scientific Workflows onto Distributed Systems. Scientific Programming, 13 (3). 219-237.
[7] Foster, I., "Compositional Parallel Programming Languages", ACM Transactions on Programming Languages and Systems 18 (4), 1996, pp. 454-476.
[8] Gray, J. "Distributed Computing Economics", Technical Report MSR-TR-2003-24, Microsoft Research, Microsoft Corporation, 2003.
[9] Hey, T., Trefethen, A., The data deluge: an e-sicence perspective, Gid Computing: Making the Global Infrastructure a Reality, 2003, Wiley.
[10] Hill, M., and Marty, M., Amdahl's Law in the Multicore Era. The 14th International Symposium on High-Performance Computer Architecture, 2008
[11] Ludäscher, B., Altintas, I., Berkley, C., Higgins, D., Jaeger-Frank, E., Jones, M., Lee, E., Tao, J., Zhao, Y., Scientific Workflow Management and the Kepler System. Concurrency and Computation: Practice & Experience, 2005.
[12] Oinn, T., Addis, M., Ferris, J., Marvin, D., Senger, M., Greenwood, M., Carver, T., Glover, K., Pocock, M. R., Wipat, A., Li, P., Taverna: A tool for the composition and enactment of bioinformatics workflows. Bioinformatics Journal, 20(17):3045–3054, 2004.
[13] The Potential of the Cell Processor for Scientific Computing. Computational Research Division, Lawrence Berkeley National Laboratory, 2007.
[14] Raicu, I., Zhao, Y., Dumitrescu, C., Foster, I., Wilde, M., "Falkon: a Fast and Lightweight Task Execution Framework", IEEE/ACM Supercomputing, 2007.
[15] Raicu, I., Zhao, Y., Foster, I., Szalay, A. "Accelerating Large-scale Data Exploration through Data Diffusion", ACM/IEEE Workshop on Data-Aware Distributed Computing, 2008.
[16] Shukla, D., Schmidt, R., Essential Windows Workflow Foundation (Microsoft .NET Development Series), Addison-Wesley Professional, 2006.
[17] Stevens, R., The LLNL/ANL/IBM Collaboration to Develop BG/P and BG/Q, DOE ASCAC Report, 2006.
[18] Sutter, H., The Free Lunch Is Over: A Fundamental Turn Toward Concurrency in Software, Dr. Dobb's Journal 20(3), March 2005.
[19] Szalay, A., Bunn, A.., Gray, J., Foster, I., Raicu, I. "The Importance of Data Locality in Distributed Computing Applications", NSF Workflow Workshop 2006.
[20] Thain, D., Tannenbaum, T., Livny, M. "Distributed Computing in Practice: The Condor Experience" Concurrency and Computation: Practice and Experience, Vol. 17, No. 2-4, pages 323-356, February-April, 2005.
[21] UltraSPARC® T2 Processor, The world's first true system on a chip, Sun Microsystems Datasheet, 2008.
[22] Zhao, Y., Hategan, M., Clifford, B., Foster, I., Laszewski, G.v., Raicu, I., Stef-Praun, T., Wilde, M., Swift: Fast, Reliable, Loosely Coupled Parallel Computation, IEEE Workshop on Scientific Workflow (SWF07), Collocated with SCC 2007.
[23] Zhao, Z., Belloum, A., de Laat, C., Adriaans, P., Hertzberger, R.: Distributed execution of aggregated multi domain workflows using an agent framework. IEEE Workshop on Scientific Workflow, 2007.